\documentclass[pre,a4paper,amsfonts,amssymb,onecolumn,noshowpacs,showkeys,nofootinbib,floats]{revtex4}

\usepackage{amsfonts}
\usepackage{amsmath}
\usepackage{amssymb}
\usepackage{graphicx}

\begin{document}

\title{Power-law distributions in economics: a nonextensive statistical approach}

\author{S\'{\i}lvio M. Duarte Queir\'{o}s}
\email[e-mail address: ]{sdqueiro@cbpf.br}
\affiliation{Centro Brasileiro de Pesquisas F\'{\i}sicas, 150, 22290-180, Rio de Janeiro - RJ, Brazil}
\author{Celia Anteneodo}
\email[e-mail address: ]{celia@cbpf.br}
\affiliation{Departamento de F\'{\i}sica, Pontif\'{\i}cia Universidade Cat\'olica do Rio de Janeiro, 
CP 38071, 22452-970, Rio de Janeiro, Brazil}
\author{Constantino Tsallis}
\email[e-mail address: ]{tsallis@santafe.edu}
\affiliation{Santa Fe Institute, 1399 Hyde Park Road, Santa Fe - NM, USA}
\affiliation{Centro Brasileiro de Pesquisas F\'{\i}sicas, 150, 22290-180, Rio de Janeiro - RJ, Brazil}

\date{\today}

\begin{abstract}
The cornerstone of Boltzmann-Gibbs ($BG$) statistical mechanics is  the Boltzmann-Gibbs-Jaynes-Shannon 
 entropy $S_{BG} \equiv -k\int dx \, f(x)\ln f(x)$, where $k$ is 
a positive constant and $f(x)$ a probability 
density function. 
This theory has exibited, along more than one century, great success in the treatment of systems where short
spatio/temporal correlations dominate.
There are, however, anomalous natural and artificial systems that violate the basic requirements for
its applicability.
Different physical entropies, other than
the standard one, appear to be necessary in order to satisfactorily deal with such anomalies. 
One of such entropies is 
$S_q \equiv k \,(1-\int dx \,[f(x)]^q)/(1-q)$ (with $S_1=S_{BG}$), where the entropic
index $q$ is a real parameter. It has been proposed as the basis for a generalization, 
referred to as {\it nonextensive statistical mechanics}, of the $BG$ theory. $S_q$ shares 
with $S_{BG}$ four remarkable properties, namely {\it concavity} ($\forall q>0$), 
{\it Lesche-stability} ($\forall q>0$), {\it finiteness of the entropy production per unit time} 
($q \in  \Re$), and {\it additivity} (for at least a compact support of $q$ including $q=1$). 
The simultaneous validity of these properties suggests that $S_q$ is appropriate for bridging, 
at a macroscopic level, with classical thermodynamics itself.  
In the same natural way that exponential probability functions arise
in the standard context,
power-law tailed distributions, even with exponents {\it out} of the L\'evy range, 
arise in the nonextensive framework. 
In this review, we intend to show that many processes of interest in economy, 
for which fat-tailed probability functions are empirically observed, 
can be described in terms of the statistical mechanisms that underly the 
nonextensive theory.  
\end{abstract}


\maketitle

\section{Introduction}

The concept of ``entropy'' (from the greek $\tau \rho o\pi 
\acute{\eta}$, transformation), was introduced in $1865$ by Rudolf Julius Emmanuel  Clausius in the 
context of Thermodynamics\cite{fermi}. This was motivated by his
studies on reversible and
irreversible transformations, as a measure of the amount of energy in a 
physical system, that cannot be used to perform work. More specifically, 
Clausius defined \textit{change in entropy }of a thermodynamic system, during 
some reversible process where a certain amount of heat $\delta Q$ is  
transported at constant temperature $T$, as $\delta S=\delta Q/T$. 
We can consider {\em entropy} as the cornerstone of 
Thermodynamics, since all the thermodynamical principles involve, directly or indirectly, 
this fundamental concept. 

The first connection between the macroscopic Clausius' entropy of a system 
and its microscopic configurations was done by Ludwig 
Boltzmann in $1877$~\cite{fermi}. 
Studying the approach to equilibrium of an ``ideal"
gas~\cite{boltz}, he
realized that the entropy could be related to the number of possible
microstates
compatible with the thermodynamic properties of the gas. For an
isolated system in its terminal stationary state ({\it thermal equilibrium}),
Boltzmann observation can be expressed as 
\begin{equation}
S_{BG}=k\,\ln W ,  \label{S_B}
\end{equation}
where $k$ is a positive constant and $W $ the number of microstates consistent
with the macroscopic state.
This famous equation, known as \textit{Boltzmann principle},  
can be regarded as a pillar in the foundations of \textit{statistical mechanics}, 
which aims to describe the thermodynamic state of a system   
through the statistical approach of its microstates.

For systems which are not isolated but instead are in contact with some 
reservoir (of heat, particles, etc.), it is
possible to derive (under some assumptions),
from Eq.~(\ref{S_B}) $\,$ \cite{SBG} , the celebrated Boltzmann-Gibbs entropy 
\begin{equation}  \label{S_BG}
S_{BG}=-k\sum\limits_{i=1}^{W}p_{i}\,\ln p_{i},
\end{equation}
where $p_i$,  subject to the normalization 
condition $\sum_{i=1}^{W} p_{i}=1$, is the probability of the 
microscopic configuration $i$.
In particular, for equiprobability, $p_{i}=1/W$, $\forall _{i}$,
then Boltzmann-Gibbs entropy (\ref{S_BG}) reduces to (\ref{S_B}). 
Since it refers to microscopic states, the Boltzmann principle should  be 
derivable from microscopic dynamics. However, the implementation of such 
calculation has not been yet achieved. Consequently, $BG$ statistical mechanics
still remains based on hypothesis such as Boltzmann's \textit{Stosszahlansatz} (molecular chaos 
hypothesis)\cite{boltz} and {\em ergodicity} \cite{khinchin}.
It can be easily shown that entropy (\ref{S_BG}) is \textit{nonnegative}, 
\textit{concave}, \textit{experimentally robust} (or
\textit{Lesche-stable})\cite{SBG,props}, and leads to a {\it finite entropy production per unit time}\cite{latorabaranger99}.
Moreover, it is \textit{additive}. In other words, if $A$ and $B$ are two 
\textit{probabilistically independent} subsystems, i.e., $p_{ij}^{A+B}=p_{i}^{A}p_{j}^{B}$,
then it is straightforwardly verified that 
$$ S_{BG}\left( A+B\right) =S_{BG}\left( A\right) +S_{BG}\left( B\right), $$
hence, if we have $N$ subsystems, $S_{BG}(N)=NS_{BG}(1)$, where the notation is self-explanatory. More generally, when correlations are ``weak" enough, $S_{BG}$ is {\em extensive}, i.e., 
such that the $\lim_{N\rightarrow \infty} S_{BG}(N)/N$ is {\it finite}.

Despite the lack of first-principle derivations, Boltzmann-Gibbs statistical 
mechanics has a history plenty of successes in the treatment of systems 
where short spatio/temporal interactions dominate. This kind of 
interactions favor ergodicity and independence properties, necessary in 
Khinchin's approach of $S_{BG}$ \cite{khinchin}. Consequently, it is perfectly plausible that   
physical entropies other than the Boltzmann-Gibbs one 
can be defined in order to suitably treat anomalous systems, for which ergodicity 
and/or independence are not verified. As examples of anomalies we can 
mention: metastable states in long-range interacting Hamiltonian systems, 
metaequilibrium states in small systems (i.e., systems whose number of 
particles is much smaller than Avogrado's number), glassy systems, some types 
of dissipative systems, systems that in some way violate ergodicity, and, \textit{last but not least}, systems with 
non-Markovian memory, like it seems to be the case of financial ones. Generically speaking, systems that may possibly have a multifractal, scale-free or 
hierarchical structure in the occupancy of their phase space.

Motivated by this scenario, one of us proposed in $1988$ the
entropy\cite{ct88}
\begin{equation}  \label{Sq}
S_{q}=k\frac{1-\sum\limits_{i=1}^{W}  p_{i} ^{q}}{1-q}\qquad
\left( q\in \Re \right) ,  \end{equation}
which generalizes $S_{BG}$, such that $\lim_{q\rightarrow 1}S_{q}=S_{BG}$, 
as the basis of a possible generalization of Boltzmann-Gibbs statistical 
mechanics\cite{further} and where the \textit{entropic index }$q$
should be determined
\textit{a priori } from microscopic dynamics. Just like $S_{BG}$, 
$S_{q}$ is \textit{nonnegative}, \textit{concave},  \textit{experimentally 
robust} (or \textit{Lesche-stable}) ($\forall  {q>0}$), and leads to a {\it finite entropy production per unit time}  (\cite{qprops} and
references therein).
Moreover, it has been  recently shown\cite{additive} that it is also  
\textit{additive}, i.e., 
$$S_{q}\left( A_{1}+A_{2}+\ldots +A_{N}\right)
=\sum\limits_{i=1}^{N}S_{q}\left( A_{i}\right),$$
for special kinds of {\it correlated} systems, more precisley when phase-space is occupied in
a scale-invariant form.

Since its proposal, entropy (\ref{Sq}) has been the source of numerous
results in both fundamental and applied physics as well as in other scientific areas
such as biology, chemistry, economics, geophysics and medicine
\cite{applications}.
It is our purpose here to review some new results concerning applications 
to economics and more specifically to the theory of finances.

\section{Probability density functions from the variational principle}

Before discussing some specific quantities of financial interest, let us see 
the form of the probability density function naturally derived from the 
variational principle related to entropy (\ref{Sq}). 
We consider its continuous version, i.e.,   
\begin{equation}
S_{q}=k\frac{1-\int \left[ p\left( x\right) \right] ^{q}\ dx}{1-q}.
\label{sq-cont}
\end{equation}
Natural constraints in the maximization of (\ref{sq-cont}) are
\[
\int p\left( x\right) \ dx=1\,,
\]
corresponding to normalization, and   
\begin{equation} \label{mean}
\int x\frac{\ \left[ p\left( x\right) \right] ^{q}}{\int \left[ p\left(
x\right) \right] ^{q}dx}\ dx\equiv \left\langle x\right\rangle _{q}=\bar{\mu}
_{q} \, ,
\end{equation}
\begin{equation} \label{variance}
\int \left( x-\bar{\mu}_{q}\right) ^{2}\frac{\ \left[ p\left( x\right) 
\right] ^{q}}{\int \left[ p\left( x\right) \right] ^{q}dx}\ dx\equiv
\left\langle \left( x-\bar{\mu}_{q}\right) ^{2}\right\rangle _{q}=\bar{\sigma
}_{q}^{2}\,,
\end{equation}
corresponding to the {\em generalized} mean and variance of a relevant 
quantity $x$, respectively.
It is noteworthy that, averages weighted with a function of the
probabilities  as in
(\ref{mean})-(\ref{variance}) allow to mimic the way individuals
behave in face to
risky choices\cite{risk} (see also \cite{also}). 
In fact, the {\em prospect} theory proposed by Kahneman and
Tversky\cite{at1} for
analyzing decision-making under risk, is founded on the concept of 
``decision weights'' that can be modeled by akin functional forms\cite{shape}.

From the variational problem for (\ref{sq-cont}) under the above constraints, 
one obtains
\begin{equation}\label{pq-1}
p\left( x\right) =\mathcal{A}_{q}\left[ 1+\left( q-1\right) \mathcal{B}
_{q}\left( x-\bar{\mu}_{q}\right) ^{2}\right] ^{\frac{1}{1-q}},\qquad \left(
q<3\right),   \end{equation}
where,
\[
\mathcal{A}_{q}=\left\{ 
\begin{array}{ccc}
\frac{\Gamma \left[ \frac{5-3q}{2-2q}\right] }{\Gamma \left[ \frac{2-q}{1-q}
\right] }\sqrt{\frac{1-q}{\pi }\mathcal{B}_{q}} & \Leftarrow  & q<1 \\[3mm] 
\frac{\Gamma \left[ \frac{1}{q-1}\right] }{\Gamma \left[ \frac{3-q}{2q-2}
\right] }\sqrt{\frac{q-1}{\pi }\mathcal{B}_{q}} & \Leftarrow  & q>1
\end{array}
\right. ,
\]
and
\[
\mathcal{B}_{q}=\left[ \left( 3-1\right) \,\bar{\sigma}_{q}^{2}\right] ^{-1}.
\]
The upper bound $q=3$ guarantees normalizability. 
Defining the $q$-{\it exponential} function as
\[
e_{q}^{x} \equiv \left[ 1+\left( 1-q\right) \,x\right] ^{
\frac{1}{1-q}}\qquad \left( e_{1}^x \equiv e^{x}\right) ,
\]
($e_q^x=0$ if $1+(1-q)x \le0$) we can rewrite the probability density (\ref{pq-1}) as 
\begin{equation}\label{pq}
p\left( x\right) =\mathcal{A}_{q}\,e_{q}^{-\mathcal{B}_{q}\left( x-\bar{\mu}%
_{q}\right) ^{2}},  \end{equation}
hereon referred to as $q$-{\em Gaussian} probability density function.

For $q=\frac{3+m}{1+m}$, the $q$-Gaussian form recovers the Student's
$t$-distribution
with $m$ degrees of freedom ($m=1,2,3,\ldots $). Consistently, for 
$q>1,$ (\ref{pq}) presents an asymptotic \textit{power-law} behavior. 
Also, if $q=\frac{n-4}{n-2}$ with $n=3,4,5,\ldots $,  
 \ $p\left( x\right) $ recovers the $r$-distribution with $n$ degrees of
freedom. Consistently, for $q<1$, $p\left(x\right)$ has a \textit{compact support} which is defined
by the condition
$\left\vert x-\bar{\mu}_{q}\right\vert \leq
\sqrt{\frac{3-q}{1-q}\,\bar{\sigma}_{q}^{2}}$.

Many other entropic forms have been introduced in the literature for various interesting purposes. One of the advantages of entropy (\ref{Sq}) is that it yields power-law tails, which play a particularly relevant role, as well known.

\begin{figure}[hbt]
\begin{center}
\includegraphics[width=0.4\columnwidth,angle=0]{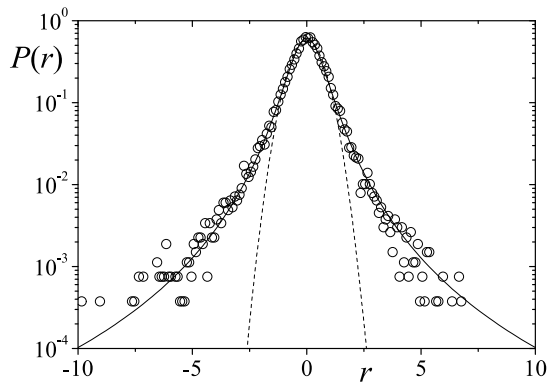}
\end{center}
\caption{
Probability density function of returns, $P(r)$, {\it versus} return, $r$. 
The symbols represent $P(r)$ for the Dow Jones Industrial daily return index 
from 1900 until 2003. The solid line represents the best q-Gaussian 
numerical adjust with $q=1.54$ and $\sigma_{q}^{2}=0.338$ (as obtained
in~\cite{smdq-quantf}) and the dashed line a Gaussian fit.}
\label{fig-dj1}
\end{figure}

Let us recall succinctly the two basic central limit theorems:
(1) A convoluted distribution with a finite second moment approaches, 
in the limit of $N\to\infty$ convolutions, a Gaussian attractor;  
(2) A convoluted distribution with a divergent second moment, approaches, 
in the same limit, a L\'evy distribution $L_\gamma(x)$ (with
$0<\gamma<2$)\cite{andre}.
However, through dynamics different from the convolution one, for
instance with some
sort of memory across successive steps (i.e., {\it nonindependence} of the successive distributions),  different ubiquitous distributions 
might arise (see also \cite{additive}). 
If the Laplacian term in the linear diffusion equation is a standard 
second derivative or a fractional derivative, the Gaussian or the 
L\'evy distributions are respectively  attained as solutions. 
However, much more complex and rich dynamics clearly exist in nature, for 
example those associated with a variety of nonlinear Fokker-Planck equations 
involving nontrivial correlations, multiplicative noise or other effects, as 
we will see in the following Section. 
Moreover, simple convolutions allow only for asymptotic behavior like the 
$q=1$ (Gaussian) or $q>5/3$ (L\'evy distributions) ones. 
But they do {\em not} allow fat-tailed distributions associated 
with $1<q\le 5/3$. However, many complex systems in nature as well as in 
social sciences exhibit exponents which precisely belong to that interval.    
An example of financial interest is exhibited in Fig~\ref{fig-dj1}. 

\section{Underlying stochastic processes}

The Gaussian distribution, recovered in the limit $q\rightarrow 1$ of
expression (\ref{pq}),
can be derived on a variety of grounds. 
For instance, it 
has been derived, through arguments based on the dynamics, by L.
Bachelier in his 1900 work on price changes in Paris stock market, and also
by A. Einstein in his 1905 article on Brownian motion. In particular, 
starting from a Langevin dynamics, one is able to write the corresponding 
Fokker-Planck equation and, from it, to obtain as solution the
Gaussian distribution.
Analogously, it is also possible, from certain classes of stochastic
differential equations
and their associated Fokker-Planck equations, 
to obtain the distribution given by (\ref{pq}). 

In this section we will discuss dynamical mechanisms which lead  
to probability functions with asymptotic power-law behavior of the
$q$-Gaussian form.

\subsection{Stochastic processes with additive multiplicative structure}
\label{Sec:additive-multiplicative}

Microscopic dynamics containing multiplicative noise may be encountered 
in many dynamical processes and, due to its significance,  
has been the subject of numerous studies in the last decades.
The presence of additive noise, besides the multiplicative contribution, 
is in fact a quite realistic ingredient since not all the fluctuations 
are processed multiplicatively. In previous work\cite{multi1}, we considered 
a class of differential stochastic equations of the form

\begin{equation}
\dot{x}\;=\;f(x) + g(x)\zeta(t)  \;+\; \eta(t)\,,
\end{equation}
where $f,g$ are arbitrary functions of the stochastic variable $x$, and 
$\zeta(t),\eta(t)$, two independent zero-mean Gaussian white noises with 
variance $M^2$ and $A^2$, respectively. 
We have shown that for special forms of the deterministic force, namely 
$f(x)=-\gamma g(x)g'(x)$, the stationary probability density functions 
(by using the It\^o prescription) are of the form 
\begin{equation} \label{ps}
P_s(x)\propto e_q^{-\beta[g(x)]^2}\, ,
\end{equation}
where $\beta \equiv (\gamma+M^2)/A^2$ and 

\begin{equation}
q\;=\;\frac{1+2M^2/\gamma}{1+M^2/\gamma} \,.
\end{equation}
From the point of view of entropy $S_q$, the density function 
(\ref{ps}) derives from the variational principle, under the 
constraints of normalization and generalized variance of $g(u)$. 

In particular, $q$-Gaussian distributions can be derived from a 
stochastic process of the linear form\cite{multi1,multi2,multi3,multi4}

\begin{equation} \label{langevin_gen}
\dot{x}\;=\;-\gamma x  +x\zeta(t)  \;+\; \eta(t)\,.
\end{equation}

In fact, its associated Fokker-Planck equation is

\begin{equation} 
\frac{\partial P}{\partial t} = \gamma\frac{\partial (xP)}{\partial x} + 
\frac{A^2}{2}\frac{\partial^2[(1+(M/A)^2x^2)P]}{\partial x^2} 
\end{equation}
that has the alternative form

\begin{equation} \label{nlfp}
\frac{\partial P}{\partial t} = \gamma\frac{\partial (xP)}{\partial x} 
+\frac{D}{2}\frac{\partial P^\nu}{\partial x^2} \,,
\end{equation}
where $\nu=2-q$, and $D$ is a constant related to the other model parameters. 
Eq. (\ref{nlfp})  is a {\em nonlinear} diffusion equation, familiarly known as 
porous media equation. Its steady state solution has the form 

\begin{equation}
P_s(x)\propto e_q^{-\beta x^2}\,
\end{equation}
with $q$ and $\beta$ defined as above. 
In the particular case $\nu=1=q$ the standard 
Gaussian steady state is obtained, corresponding to a purely additive process. 

Taking into account that empirical returns where found to follow a 
$q$-Gaussian distribution\cite{lisa} (see also Fig.~\ref{fig-dj1}), 
Eq. (\ref{langevin_gen}), complemented by the 
It\^o prescription, provides a simple mechanism to model the dynamics
of prices.
Along similar lines it has been worked out,   
for instance, an option-pricing model which is more realistic
than the celebrated Black-Scholes one\cite{lisa} (recovered as the $q=1$ particular case).

The $q$-exponential character of the solutions of Eq.~(\ref{nlfp}) is
not exclusive of
the steady state but it also emerges along the time
evolution\cite{multi1,FP_nl}.
In the presence of multiplicative noise, the system variables directly
couple to noise.
Therefore, behaviors are observed that can not occur in the presence of 
additive noise alone. On the other hand the additive noise plays a fundamental 
role allowing the existence of a reasonable and normalizable steady state 
by avoiding collapse of the distribution at the origin. 
The particular interplay between additive and multiplicative noises,
as well as 
that between deterministic and stochastic drifts, can lead to the
appearance of $q$-exponential forms.

The $q$-exponential distributions include the Boltzmann-Gibbs one as a
special case ($q=1$).
While the latter corresponds to the standard thermal equilibrium, the
$q\neq1$ is expected to
be related to quasi-stationary or long-living out-of-equilibrium regimes.   

\subsection{Stochastic processes with varying intensive parameters}
\label{Sec:superstat}

Intricate dynamical behavior is a common feature of many
non-equilibrium systems,
which can be also characterized by power-law probability density functions. 
To this class belong systems whose dynamical behavior shows
spatio/temporal fluctuations of some intensive
quantity. 
This quantity may be the inverse temperature, like in the case of
interactions of hadrons from cosmic
rays in emulsion chambers; the energy dissipation in turbulent fluids, 
or simply the width of some white noise,
as assumed in many financial models, such as in the famous Heston
model\cite{heston}.
The connection between 
this sort of dynamics and nonextensive entropy was first made by G. Wilk and
Z. W\l odarczyk \cite{wilk}
and later extended by C. Beck and E.G.D. Cohen\cite{beck-cohen}, 
who called it {\it superstatistics}. 
In this ``statistics of statistics", Beck and Cohen aimed to 
treat non-equilibrium systems from the point of view of long-living
stationary states
characterized by a temporally or spatially fluctuating
intensive parameter. Such condition can be mathematically expressed by
\begin{equation}
B[E(z)]=\int_{0}^{\infty }f\left( \beta \right) \,e^{-\beta E(z)}\,d\beta
\label{superstatistics}
\end{equation}
where $B[E(z)]$ is a kind of effective Boltzmann factor, $E(z)$ a function
of some relevant variable $z$, and $f(\beta )$ the probability density function of the 
inverse temperature $\beta$. Superstatistics is intimately connected with
nonextensive statistical mechanics. More precisely, it is possible to derive a
generalized Boltzmann
factor which is exactly $B[E]$, when $f(\beta )$ is the Gamma distribution,
i.e., 
\[
e_{q}^{-\beta ^{\prime }E(z)}=\int \frac{e^{-\beta /b}}{b\,\Gamma \left[ c%
\right] }\left( \frac{\beta }{c}\right) ^{c-1}e^{-\beta /b}\, , 
\]
where the  $q$-exponential functional form of the effective Boltzmann factor 
turns clearly visible its 
asymptotic power-law behavior.
It is noteworthy that the above effective Boltzmann factor is also a
good approximation
for other $f(\beta )$ probability density functions~\cite{beck-cohen}.

\section{Applications to financial observables}

\subsection{ARCH$\left( 1\right) $ and GARCH$\left(
1,1\right) $ processes from a nonextensive perspective}

The fluctuating character of volatility in financial markets has been
considered, since a few decades ago, as major responsible for price 
change dynamics \cite{mandelbrot}. 
In fact, the intermittent character of return time
series is usually associated with localized bursts in volatility and thus
called \textit{volatility clustering \cite{lo}}. The
temporal evolution of the second-order moment, known as 
\textit{heteroskedasticity \cite{engle}}, has proven to be of extreme 
importance in order to define better performing option-price 
models \cite{heston,hull,fouque} 

The first proposal aiming to modelize and analyze economical time series with
time-varying volatility was made by R.F. Engle \cite{engle}, who defined the
{\it autoregressive conditional heteroskedasticity} ($ARCH$) process. In his
seminal article, Engle stated that a heteroskedastic observable $z$ (e.g., the 
\textit{return}) would be defined as
\begin{equation}
z_{t}=\sigma _{t}\,\omega _{t},  \label{hetero-def}
\end{equation}
where $\omega _{t}$ represents an independent and identically distributed
stochastic process with null mean and unitary variance ($\left\langle \omega
_{t}\right\rangle =0$, $\left\langle \omega _{t}^{2}\right\rangle =1$)
associated to a probability density function $P_{n}\left( \omega \right) $, 
and $\sigma _{t}$ the volatility. In the same work, Engle also suggested a
simple dynamics for volatilities, a linear dependence of $\sigma _{t}^{2}$
on the $n$ previous values of $\left[ z_{t}^{2}\right] $,
\begin{equation}
\sigma _{t}^{2}=a+\sum\limits_{i=1}^{n}b_{i}\,z_{t-i}^{2},\qquad \left(
a,b_{i}>0\right) ,  \label{arch-def}
\end{equation}
afterwards named as $ARCH(n)$ \textit{linear process \cite{boller-chou}}.
The $ARCH(n)$ process is uncorrelated and for $n=1$ it presents a volatility
autocorrelation function with a characteristic time of order $\left\vert \log
b_{1}\right\vert ^{-1}$ \cite{embrechts}.

In order to give a more flexible structure to the functional form of 
$\sigma_{t}^{2}$, T. Bollerslev generalized Eq.~(\ref{arch-def}) defining the 
$GARCH(n,m)$ process\cite{boller-garch}
\begin{equation}
\sigma_{t}^{2}=a+\sum\limits_{i=1}^{n}b_{i}\,z_{t-i}^{2}+\sum%
\limits_{i=1}^{s}c_{i}\,\sigma _{t-i}^{2},\qquad \left( c_{i}>0\right) ,
\label{garch-def}
\end{equation}
which reduces to $ARCH(n)$ process, when $c_{i}=0,\ \forall _{i}$.

For the $GARCH(1,1)$ ($b_{1}\equiv b$ and $c_{1}\equiv c$), we can
straightforwardly determine the $k^{th}$ moment of the 
\textit{stationary} probability density function $P\left( z\right)$, 
particularly its second moment
\[
\bar{\sigma}^{2}\equiv \left\langle z_{t}^{2}\right\rangle =\left\langle
\sigma _{t}^{2}\right\rangle =\frac{a}{1-b-c},\qquad \left( b+c\right) <1, 
\]
and the fourth moment,  which equals
the kurtosis $( k_{x}\equiv \frac{\left\langle x^{4}\right\rangle }{%
\left\langle x^{2}\right\rangle ^{2}}) $, 
\[
\left\langle z_{t}^{4}\right\rangle =k_{z}=k_{\omega }\left( 1+b^{2}\frac{%
k_{\omega }-1}{1-c^{2}-2bc-b^{2}k_{\omega }}\right) ,\qquad \left(
c^{2}+2bc+b^{2}k_{\omega }<1\right) , 
\]
for processes with unitary variance, i.e., $\bar{\sigma}^{2}=1$.

Continuous approaches are becoming more often used (mainly in the
treatment of high-frequency data: see, e.g., \cite{fouque,scalas}).
Moreover,
$ARCH$-like processes fail in reproducing the volatility
autocorrelation function power
law behavior\cite{engle-russel}. But, despite these facts, the $ARCH$ family of 
processes (particularly $ARCH\left( 1\right) $ and $GARCH\left( 1,1\right) $) 
is still considered a cornerstone in econometrics due to its simplicity 
and satisfactory efficiency in financial time series mimicry.

Having a glance at Eq.~(\ref{hetero-def}), we can verify that the
distribution $P(z)$ of the
stochastic variable $z$ has, at each time step $t$, the same functional form of the noise
distribution, $P(\omega)$, but with a variance $\sigma _{t}$. 
This property allows one to 
look at process $\{z\}$ as a process similar to those occuring in some 
non-equilibrium systems with a longstanding stationary state.
Specifically, this principle has allowed to establish, firstly for $ARCH(1) $ 
\cite{smdq-ct-arch} and then for $GARCH(1,1) $ \cite{smdq-ct-garch}, 
a connection between $b$ and $c$, $P(z) $ and,  
$P_{n}\left( \omega \right) $, the latter assumed to be of the following
$q_{n}$-Gaussian form
\begin{equation}
P_{n}\left( \omega \right) =\frac{A_{q_{n}}}{\left[ 1+\frac{q_{n}-1}{5-3q_{n}%
}\omega ^{2}\right] ^{\frac{1}{q_{n}-1}}}
=A_{q_n}\,e_{q_n}^{-\omega^2/(5-3q_n)}
,\qquad \left( q_{n}<\frac{5}{3}%
\right) ,  \label{pqn}
\end{equation}

\begin{figure}[bht]
\begin{center}
\includegraphics[width=0.4\columnwidth,angle=0]{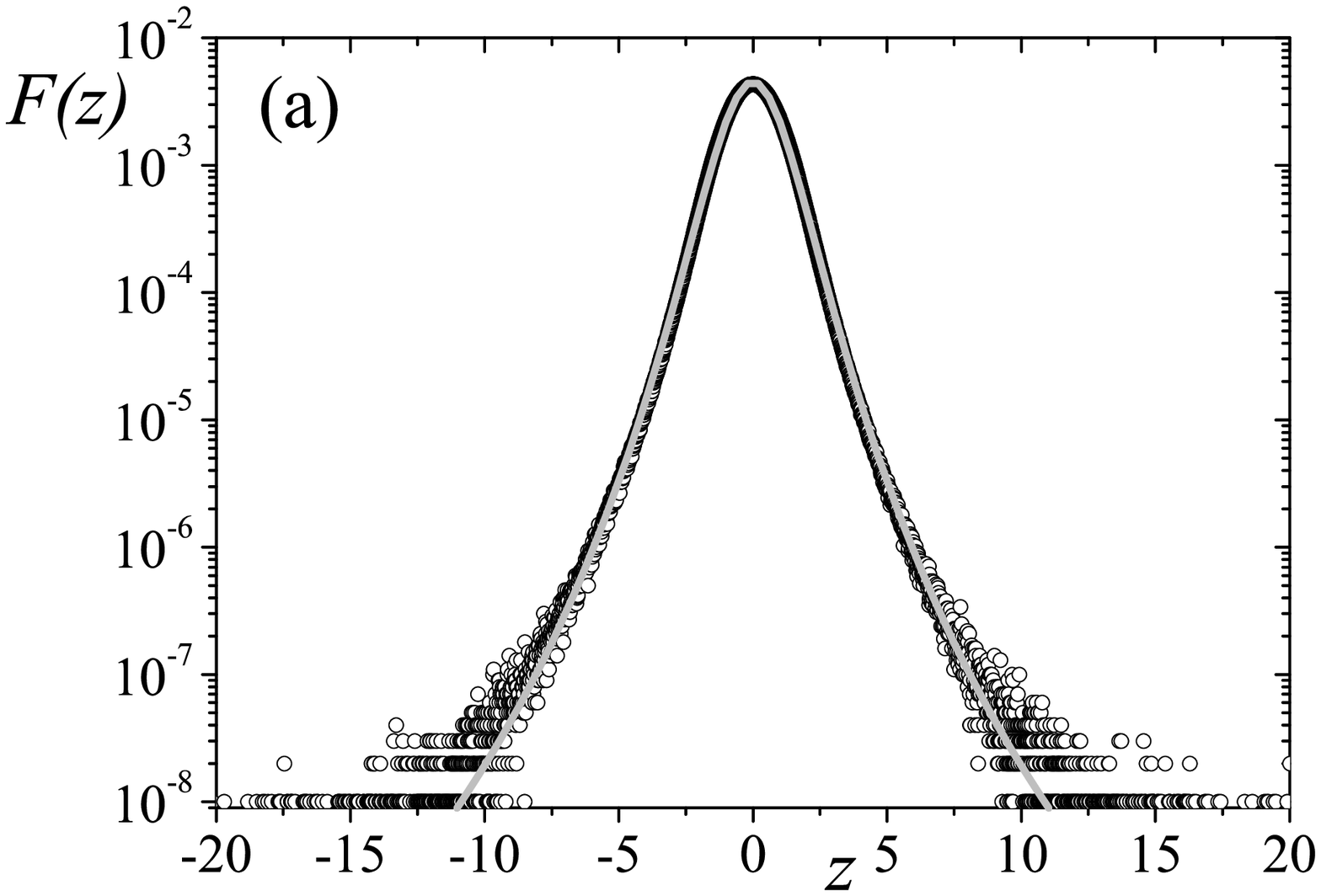}
\includegraphics[width=0.4\columnwidth,angle=0]{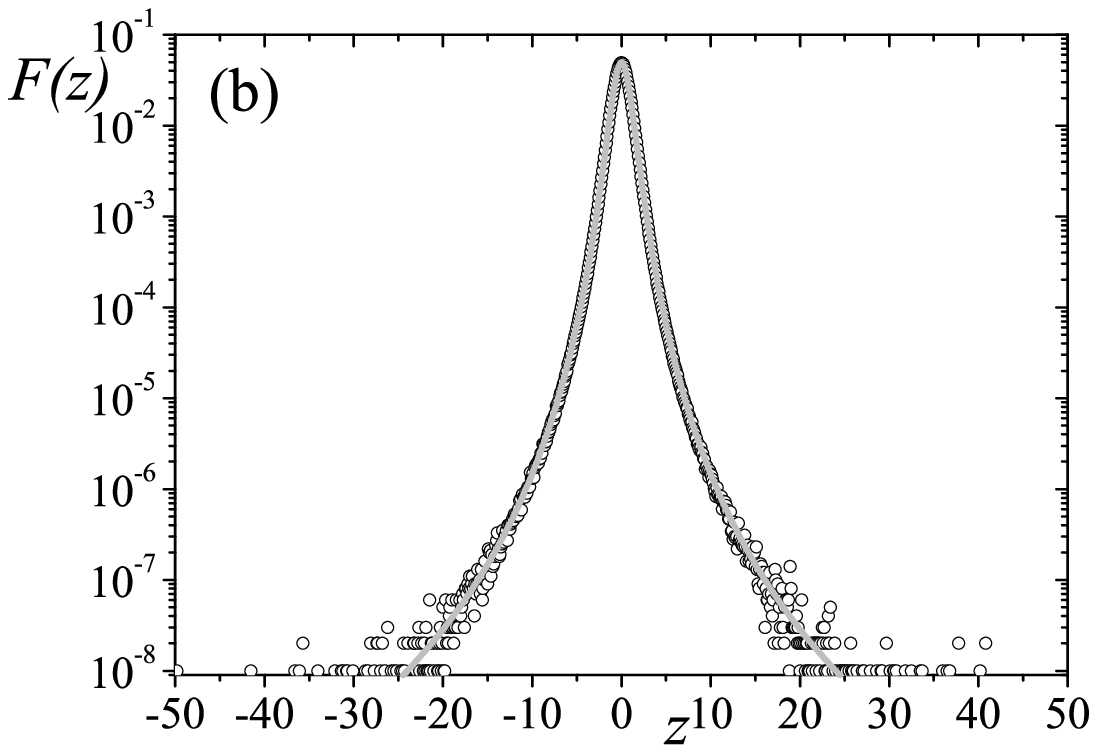}
\includegraphics[width=0.4\columnwidth,angle=0]{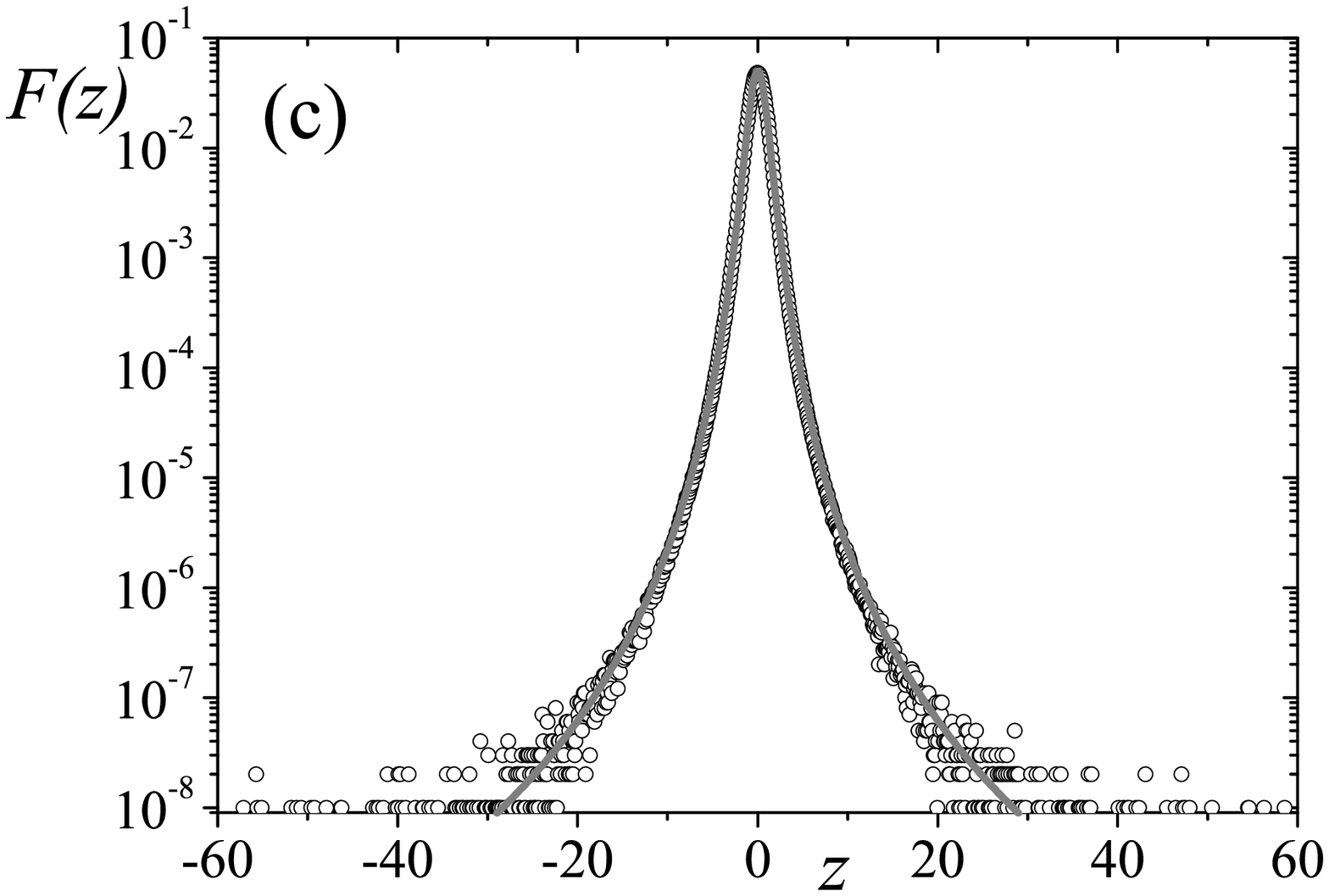}
\includegraphics[width=0.4\columnwidth,angle=0]{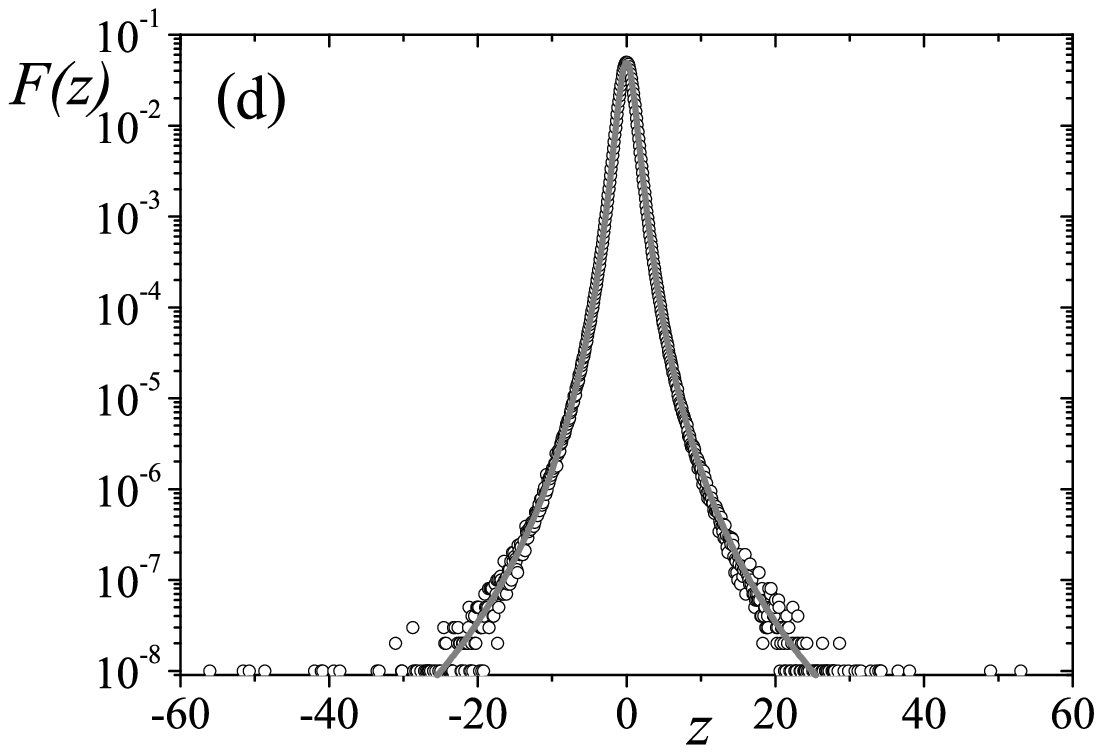}
\end{center}
\caption{Typical runs of $ARCH\left( 1\right) $ (Upper plots) and $GARCH\left(
1,1\right) $ (Lower plots) with $\bar{\sigma}^{2}=1$. For each plot the
symbols represent the relative frequency, $F\left( z\right) $, and the line
the corresponding probability function, $\int_{z-\delta }^{z+\delta 
}p\left( \tilde{z}\right) \,d\tilde{z}$. In (a) $q=1.2424$ $\left(
b=0.4;q_{n}=1\right) $; (b) $q=1.3464$ $\left( b=0.4;q_{n}=1.2\right) $; (c)
$q=1.38$ $\left( b=0.4;c=0.4;q_{n}=1\right) $; (d) $q=1.35$ $\left(
b=0.3;c=0.45;q_{n}=1.2\right) $.}
\label{fig-arch}
\end{figure}

By making the ansatz $P\left( z\right) $ $\simeq p\left( z\right) $, where 
$p\left( z\right) $ is the $q$-Gaussian probability density function which
maximizes $S_{q}$ (Eq.~(\ref{Sq})), and
by imposing the matching of second ($\bar{\sigma}^{2}=1$) and 
fourth order moments, it is possible to establish, for $GARCH\left(
1,1\right)$,
a relation containing the dynamical parameters $b$ and $c$ and
entropic indexes $q$
and $q_{n}$ : 
\begin{equation}\label{garch-rel}
b(5-3q_{n})(2-q)= 
\sqrt{(q-q_{n})-\left[(5-3\,q_{n})(2-q)
-c^{2}(5-3\,q)(2-q_{n})\right]}-c(q-q_{n}).  
\end{equation}
For $c=0$, Eq.~(\ref{garch-rel}) reduces to the one corresponding to
$ARCH\left( 1\right) $, 
\begin{equation}
q=\frac{q_{n}+2b^{2}\left( 5-3q_{n}\right) }{1+b^{2}\left( 5-3q_{n}\right) }
\label{arch-rel}
\end{equation}
and, for $b=c=0$, one has $q=q_{n}$. The validity of Eqs.~(\ref{garch-rel}) and 
(\ref{arch-rel}) is depicted in Fig.~\ref{fig-arch}. 
The discrepancy between $p\left( z\right) $ and $P\left( z\right) $
can be evaluated by computing
the sixth-order moment percentual difference, which is never greater than 
$3\%$ \cite{smdq-ct-arch,smdq-ct-garch}.

\begin{figure}[tbp]
\begin{center}
\includegraphics[width=0.5\columnwidth,angle=0]{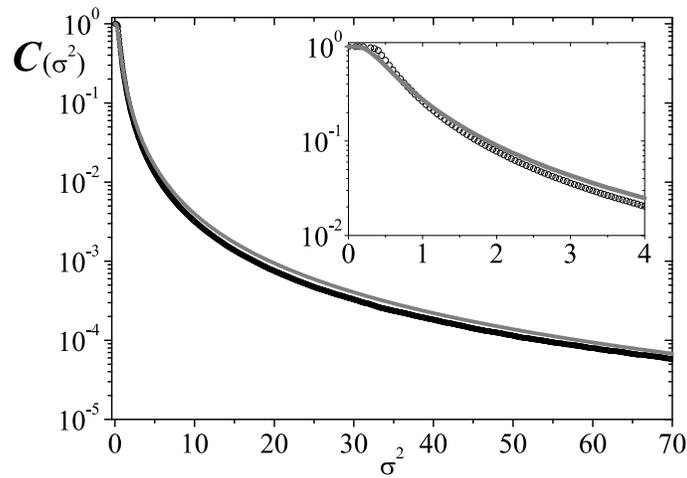}
\end{center}
\caption{The symbols in black represent the inverse cumulative
frequency, $C\left(
\sigma ^{2}\right) $, numerically obtained for a Gaussian noise with $%
b=c=0.4 $ and the gray line the respective inverse cumulative distribution, $%
\int_{\sigma ^{2}}^{\infty }P_{\sigma }\left( \tilde{\sigma}^{2}\right)
\,d\left( \tilde{\sigma}^{2}\right) $ with $\left( \kappa ,\lambda ,\bar{%
\sigma}^{2}\right) =\left( 0.444,2.125,1\right) $ for $P_{\sigma }\left(  
\tilde{\sigma}^{2}\right) $.}
\label{fig-vol}
\end{figure}

Since $\omega _{t}=z_{t}/\sigma _{t}$ and $\left\langle \omega _{t}\,\sigma
_{t}\right\rangle =0$, for $q_{n}=1$ we can write
\[
p\left( z|\sigma ^{2}\right) =\frac{1}{\sqrt{2\,\pi \,\sigma ^{2}}}
\;e^{ -\frac{z^{2}}{2 \sigma^{2}} } , 
\]
as the conditional probability density function of $z$ given $\sigma ^{2}$.
Considering that,
\[
p\left( z\right) =\int\nolimits_{0}^{\infty }p\left( z|\sigma ^{2}\right)
\,P_{\sigma }\left( \sigma ^{2}\right) \,d\left( \sigma ^{2}\right) , 
\]
and $P\left( z\right) $ $\simeq p\left( z\right) $, we obtain the 
stationary probability density function for squared
volatility\cite{smdq-ct-garch} ,
\[
P_{\sigma }\left( \sigma ^{2}\right) =\frac{\left( \sigma ^{2}\right)
^{-1-\lambda }}{\left( 2\,\kappa \right) ^{\lambda }\,\Gamma \left[ \lambda %
\right] } \;e^{ - \frac{1}{2 \kappa \sigma^{2}} } , 
\]
where 
\[
\lambda =\frac{1}{q-1}-\frac{1}{2},\qquad \kappa =\frac{1-q}{\bar{\sigma}%
^{2}\left( 3q-5\right) }. 
\]
As one can observe in Fig.~\ref{fig-vol}, the ansatz gives a quite satisfactory
description for $\sigma ^{2}$ probability density function, suggesting a
connection between the $ARCH$ class of processes and nonextensive entropy. 
These explicit formulas can be helpful in applications related, among others, to
option prices, where volatility forecasting plays a particularly important
role.

\begin{figure}[tbp]
\begin{center}
\includegraphics[width=0.5\columnwidth,angle=0]{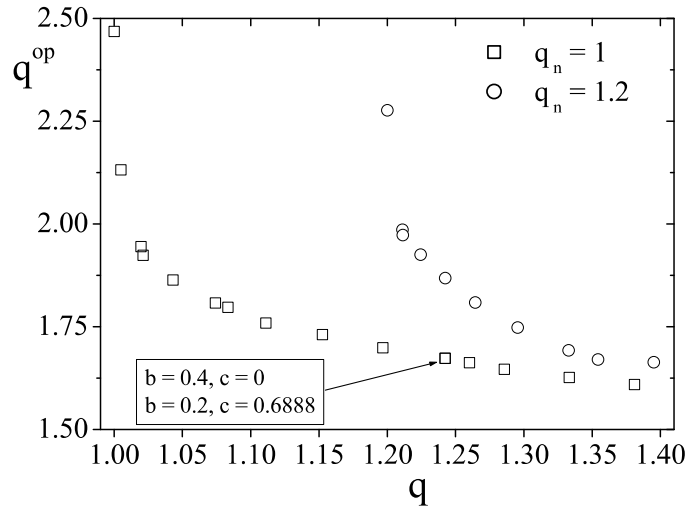}
\end{center}
\caption{Plot of $q^{op}$ \textit{versus }$q$ for typical $\left(
b,c,q_{n}\right) $\
triplets. The arrow points two examples which were obtained from 
\textit{different} triplets and nevertheless coincide in what concerns
the resulting
point $\left( q,q^{op}\right) $.}
\label{fig-qopq}
\end{figure}

Albeit uncorrelated, stochastic variables $\left\{ z_{t}\right\} $ are 
\textit{not} independent. Applying the $q$-generalized
Kullback-Leibler relative
entropy~\cite{tsallis-kl,tsallis-borland-plastino} to stationary joint
probability
density function $p_{1}\left(
z_{t},z_{t-1}\right) $ and $p_{2}\left( z_{t},z_{t-1}\right) \equiv p\left(z_t\right)p\left(z_{t-1}\right) =$ $\left[
p\left( z\right) \right] ^{2}$ it is possible to quantify the degree of
dependence between successive returns, through an optimal entropic index, $%
q^{op}$. In Ref. \cite{smdq-ct-garch}, it was verified the existence
of a direct relation between
dependence, $q^{op}$, the non-Gaussianity, $q$, and the nature of the noise, 
$q_{n}$. An interesting property emerged, namely that, whatever be the
pair $(b,c)$
that results in a certain $q$ for the stationary probability density
function, one obtains the same $q^{op}$ and, consequently, the time series
will present the same degree of dependence \cite{smdq-ct-garch}. See
Fig.~\ref{fig-qopq}.

\begin{figure}[tbp]
\begin{center}
\includegraphics[width=0.4\columnwidth,angle=0]{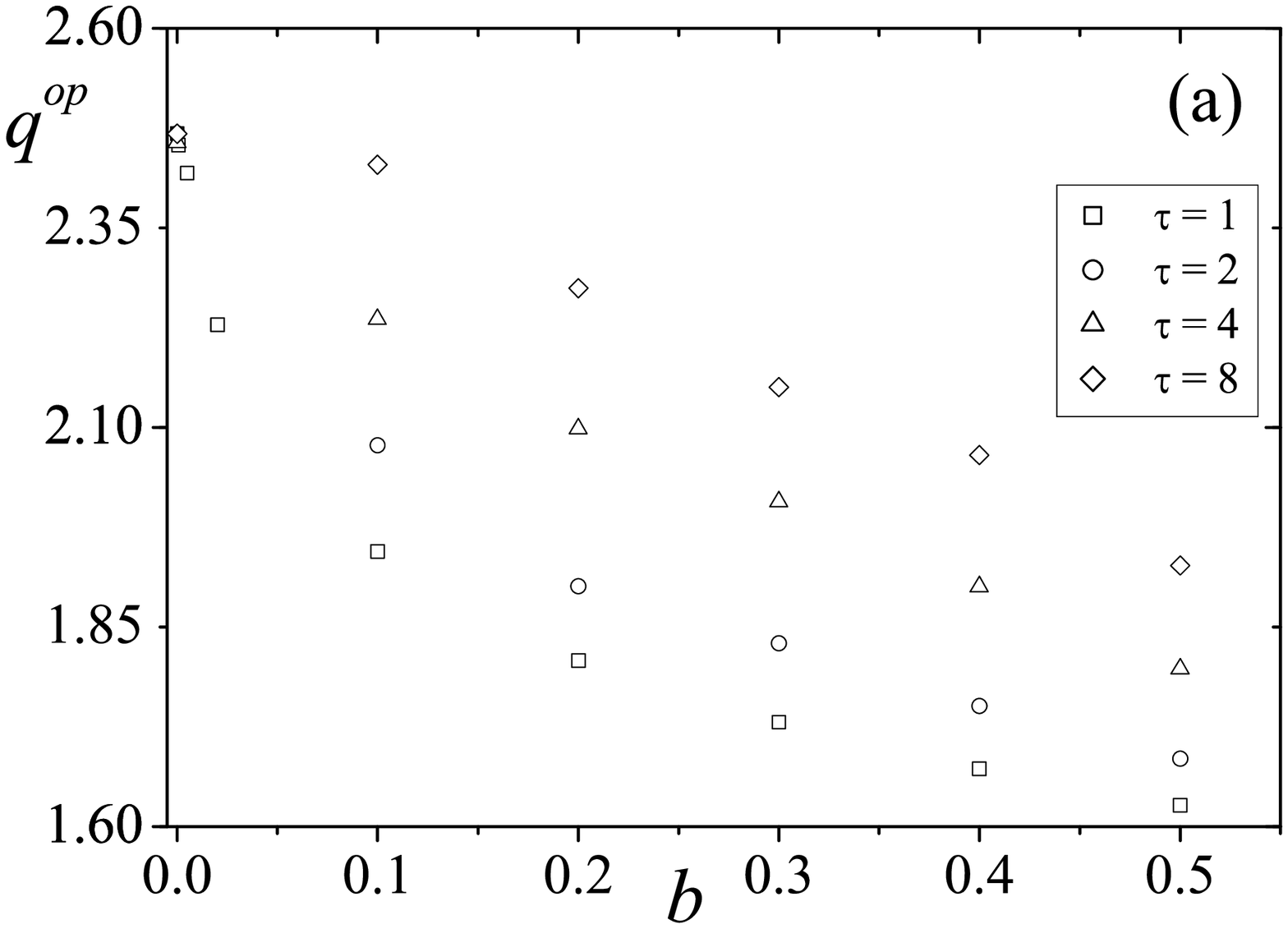}
\includegraphics[width=0.4\columnwidth,angle=0]{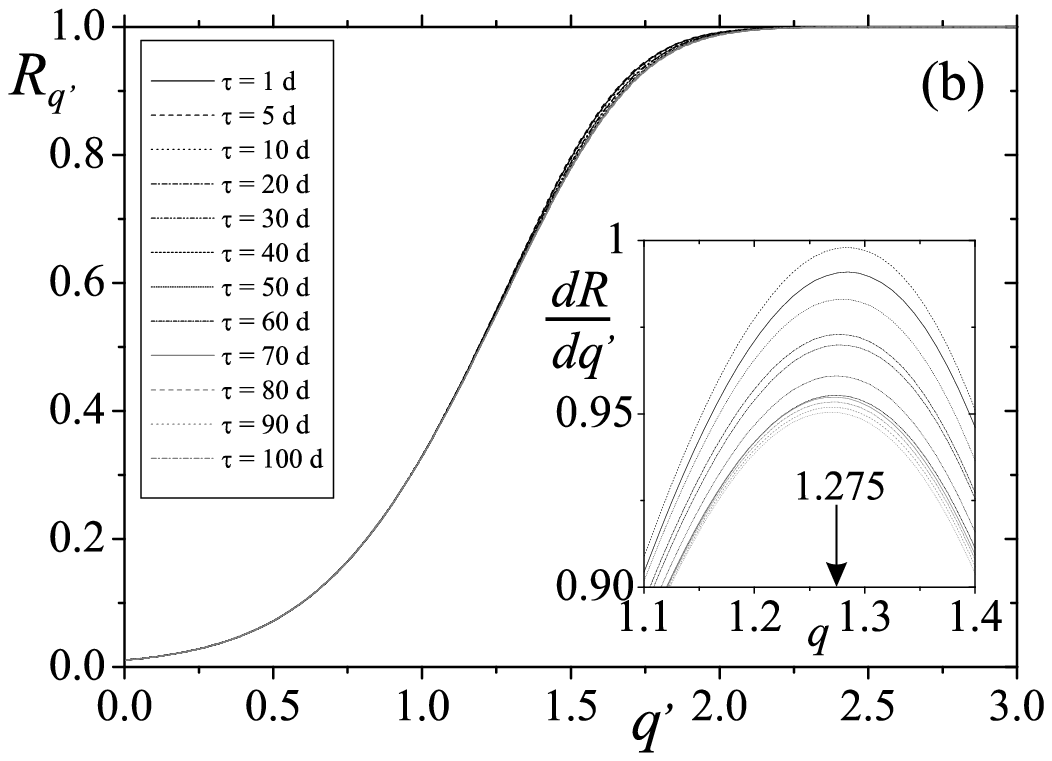}
\end{center}
\caption{In panel (a) $q^{op}$ \textit{versus} $b$ for $ARCH\left( 1\right) $
process. Even for large allowed $b$ values, the decrease of dependence  
degree  (i.e., increase of $q^{op}$) is visible when the time lag increases, which is   
not compatible with the empirical evidence of dependence degree around 
$q^{op}=1.275$ verified in Dow Jones return time series for time lags up to 
100 days, panel (b). Comparing the two figures (a)\ and (b), it appears  
that financial markets present long-memory in the volatility similar to a 
$ARCH\left( n\right) $ process with $n\gg 1$.}
\label{fig-quantf}
\end{figure}

It was also verified (see Fig.~\ref{fig-quantf}) for $ARCH\left( 1\right) $ that the degree of
dependence varies visibly with $b$ and with the lag $\tau$ between returns. This dependence 
would be related to a short-memory in volatility \cite{smdq-trends}. The
variance between these results and the empirical evidence of persistence
in the real return time series dependence degree for time intervals up to 
$100$ days recently found, shows that financial markets dynamics are, in
fact, ruled by some form of long-memory processes in volatility \cite{smdq-quantf}.

Let us comment at this point that the connection between several entropic indexes, each one related
to a different observable, is compatible with the dynamical scenario within
which the nonextensive statistical mechanics is formulated. In fact, several
entropic indexes emerge therein, coalescing onto the same value $q=1$ when
ergodicity is verified.

\newpage
\subsection{Mesoscopic models for traded volume in stock markets}

Another important observable in financial markets is 
the traded volume $V$ (the number of shares of a stock traded in a period of
time $T$). In Ref. \cite{osorio-borland-tsallis} it was proposed an
ansatz for the
probability density function of high-frequency traded volume, which
presents two power-law regimes (small and large values of $V$),
\begin{equation}
P\left( v\right) =\frac{1}{Z}\left( \frac{v}{^{\theta }}\right) ^{\alpha
}\exp _{q}\left( -\frac{v}{\theta }\right) ,  \label{prob-obt}
\end{equation}
where $v$ represents the traded volume expressed in its mean value unit  
$\left\langle V\right\rangle $, i.e., $v=V/\langle V\rangle$,  $\alpha$
and $\theta$ are positive parameters and 
$Z=\int_{0}^{\infty }(v/\theta)^{-\alpha }\exp _{q}\left(
-\frac{v}{\theta }\right) dv$.

The probability density function (\ref{prob-obt}) was recently obtained from a
mesoscopic dynamical scenario \cite{smdq-vol} based in the Feller process 
\cite{feller} (using It\^{o} definition),
\begin{equation}
dv=-\gamma \,\left( v-\frac{\alpha +1}{\beta }\right) \,dt+\sqrt{2\,v\,\frac{ 
\gamma }{\beta }}\,dW_{t},  \label{eq-vol1}
\end{equation}
where instead of being constant in time, $\beta $ varies on a time
scale much larger than the time scale of order $\gamma ^{-1}$ needed 
to reach stationarity. The deterministic term of Eq.~(\ref 
{eq-vol1}) represents a restoring market mechanism which tries to keep the
traded volume in some typical value $\Theta =\frac{\alpha +1}{\beta }$ and the
second term reflects stochastic memory and, basically, the effect of large
traded volumes. In fact, large values of $v$ will provoke large amplitude of  
the stochastic term, leading to an increase or decrease of the traded volume 
(stirred or hush stock) depending on the sign of  $W_{t}$. The
fluctuation of $\beta$,
alike to fluctuations in the mean value of $v$, can be related with changes 
in the activity volume due to agents herding mechanism caused by price
movements or news.

\begin{figure}[htb]
\begin{center}
\includegraphics[width=0.42\columnwidth,angle=0]{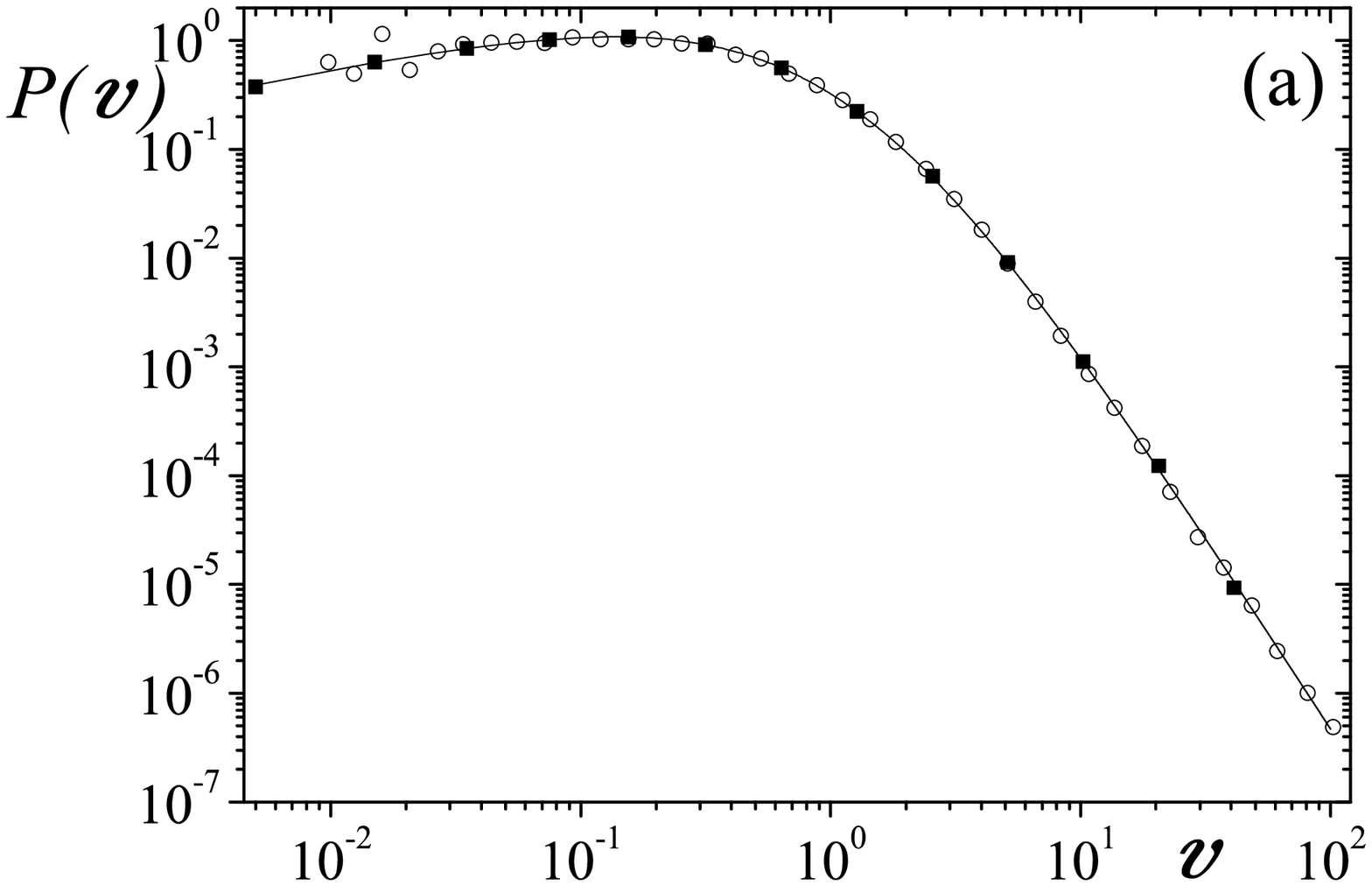}
\includegraphics[width=0.4\columnwidth,angle=0]{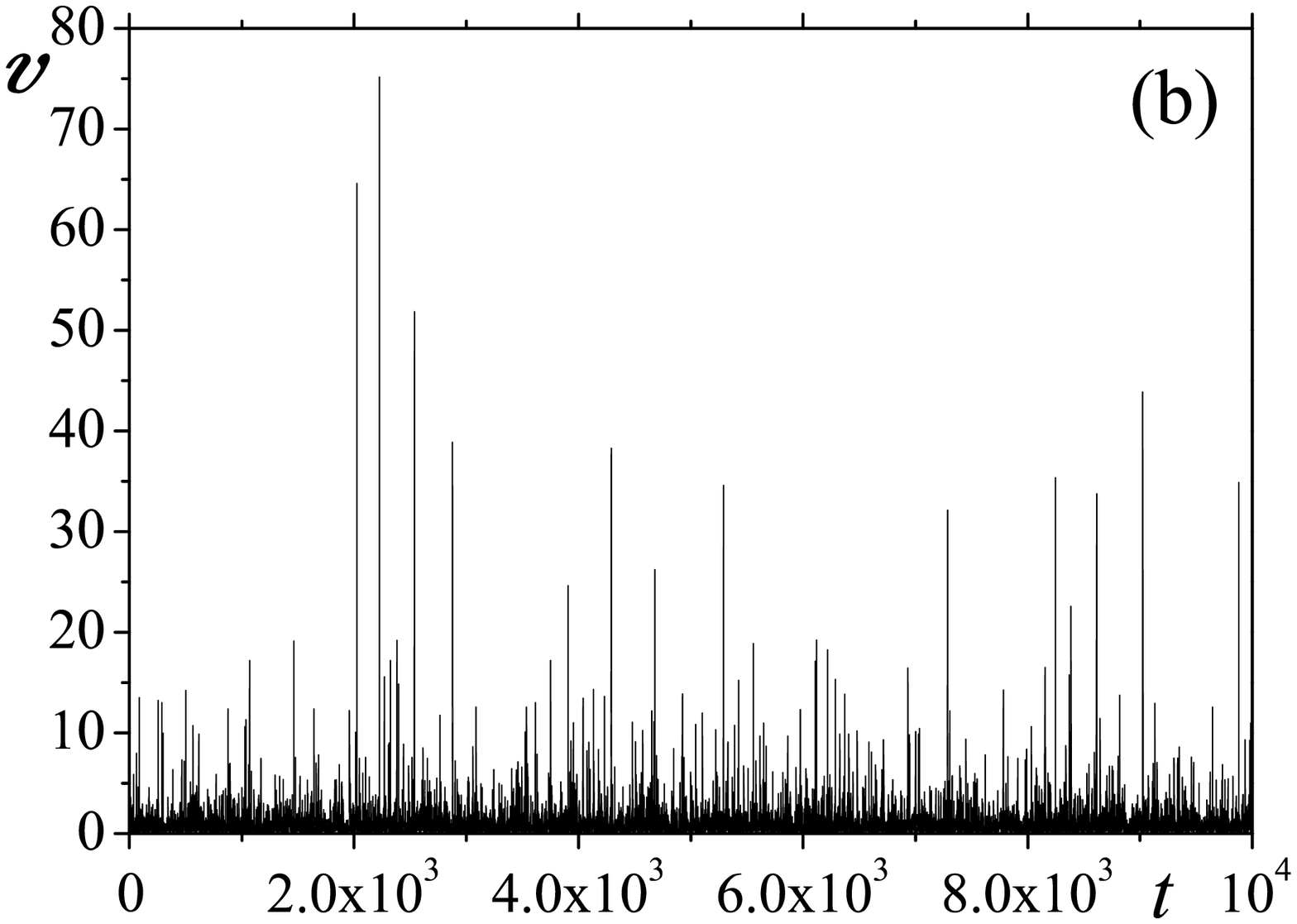}
\end{center}
\caption{In panel (a) open symbols represents the PDF for the ten-high
1 minute traded volume stocks in
NYSE exchange; solid symbols represent the PDF obtained for the
numerical realization
depicted in panel (b) and line the theoretical PDF Eq.~(\ref{pvfinal}). 
Parameters are $q =1.17$, $\alpha = 1.79$, $\lambda = 1.42$ and
$\delta = 3.09$.}
\label{fig-ny}
\end{figure}

Solving the corresponding Fokker-Planck equation for Eq.~(\ref{eq-vol1}) we
got the conditional probability of $v$ given $\beta$,
\begin{equation}
p\left( v|\beta \right) =\frac{\beta }{\Gamma \left[ 1+\alpha \right] }
\left( \beta \,v\right) ^{\alpha }\,\exp \left( -\beta \,v\right) \qquad
(\alpha >-1,\ \beta >0).  \label{pvb}
\end{equation}
Since 
\begin{equation}
P\left( v\right) =\int_{0}^{\infty }P\left( v,\beta \right) \,\,d\beta
=\int_{0}^{\infty }p\left( v|\beta \right) \,P\left( \beta \right) \,d\beta ,
\label{super}
\end{equation}
considering that $\beta $ follows a Gamma distribution,
\begin{equation}
P\left( \beta \right) =\frac{1}{\lambda \,\Gamma \left[ \delta \right] }%
\left( \frac{\beta }{\lambda }\right) ^{\delta -1}\exp \left( -\frac{\beta }{%
\lambda }\right) \qquad (\delta >0,\ \lambda >0),  \label{pb}
\end{equation}
Eq.~(\ref{super}) yields,
\begin{equation}
P\left( v\right) \equiv \frac{1}{Z}\left( \frac{v}{\theta }\right) ^{\alpha
}\exp _{q}\left( -\frac{v}{\theta }\right) ,  \label{pvfinal}
\end{equation}
where, $q=1+1/(1+\alpha +\delta)$ and $\theta =(q-1)\lambda$.
A numerical simulation for $1$ minute traded volume in NYSE stock index is
exhibited in Fig.~\ref{fig-ny}.

Another possible mechanism to describe the dynamics of volumes, has
been recently
proposed\cite{adm} on the basis of mean-reverting processes with 
additive-multiplicative structure, namely, 
\begin{equation}\label{eq-vol2}
dv=-\gamma \,\left(
v-\theta\right)\frac{1}{v}\,dt+\mu\sqrt{v}\,dW_1+\alpha dW_2,
\end{equation}
where, $\alpha$, $\mu$ are positive constants and $W_1$, $W_2$ independent 
standard Wienner processes. That is, 
following the ideas presented in Section~\ref{Sec:additive-multiplicative}, 
in addition to fluctuations endogenously processed 
by the market, other fluctuations are taken into account that affect the 
dynamics directly. 
The stationary solution of the Fokker-Planck equation associated to the 
It\^o-Langevin Eq. (\ref{eq-vol2}) has the form (\ref{pvfinal}).
Moreover, the underlying sequences present intermittent bursts
(similar to those exhibited
in Fig.~\ref{fig-ny}.b) and, preliminary results 
indicate the presence of persistent power-law correlations, as
observed in real
data sequences.

Eq. (\ref{eq-vol2}) belongs to a larger class of processes 
with two-fold power-law behavior 
that can be also suitable for volumes, as well as, for 
other mean-reverting variables such as volatilities. 

\section{Final remarks}

Additive-multiplicative processes are at the core of nonextensive statistical mechanics. 
In the same natural way that standard Brownian motion leads to Gaussianity, 
linear additive-multiplicative stochastic 
processes lead to $q$-Gaussian distributions. 
Special (scale-invariant) correlations, that forbid convergence to the usual Gauss 
or L\'evy limits, lead to a new type of statistical distributions. 
A remarkable feature is that the resulting power-law distributions may have 
exponents out of the L\'evy range, thus allowing to embrace a larger variety 
of empirical processes. 
The presence of two Gaussian white noises, one either enhanced or
reduced by internal
information, and another purely exogenous, represents quite realistic
features present in a
variety of systems, thus justifying the ubiquity of the probability
distributions associated to
such kind of processes.  
In particular, as we have shown, they allow to model the dynamics of
prizes, volatilities,
stock-volumes and other relevant financial observables. 

Another expression of the mechanism underlying the nonextensive theory 
is connected to the existence of a fluctuating intensive parameter (or
``inverse temperature'')
following the ideas that foundate the Beck and Cohen superstatistics \cite{beck-cohen}. 
We have shown that these principles allow an alternative description of the 
dynamics of stock-volumes. 
Furthermore, such kind of mechanism allows an interesting perspective
for treating the family of
{\em G/ARCH} processes. The fact that the resulting probability 
density functions can be described in terms of $q$-Gaussian distributions, 
provides a tractable way of dealing with empirical distributions that match 
the {\em G/ARCH} types. Some of the discussions presented in this review have been done at a mesoscopic scale. The determination, from more microscopic models, of the parameters used at the mesoscopic scale is certainly welcome.


\end{document}